\documentclass[prd,superscriptaddress,preprint,showkeys]{revtex4}
\usepackage{amsmath,amssymb,graphics,epsfig}
\usepackage{enumerate}
\usepackage{theorem}

\begin{document}

\title{ Covariant kinematics and gravitational bounce in Finsler space-times }

\author{A.P.Kouretsis }
\email{ akouretsis@astro.auth.gr}
\thanks{Corresponding author}
\affiliation{Section of Astrophysics, Astronomy and Mechanics, Department of Physics  Aristotle University of Thessaloniki, Thessaloniki 54124, Greece}

\author{M.Stathakopoulos}
\email{michaelshw@yahoo.co.uk}
\noaffiliation

\author{P.C.Stavrinos}
\email{pstavrin@math.uoa.gr}
\affiliation{Department of Mathematics, University of Athens, Athens 15784, Greece}

\begin{abstract}
The similarity between Finsler and Riemann geometry is an intriguing starting point to extend general relativity.
The lack of quadratic restriction  over the line element (color) naturally generalize the Riemannian case and breaks the local symmetries of general relativity. In addition, the Finsler manifold is enriched with new geometric entities and all the classical identities are suitably extended. We investigate the covariant kinematics of a medium formed by a time-like congruence. After a brief view in the general case we impose particular geometric restrictions to get some analytic insight. Central role to our analysis plays the Lie derivative where even in case of irrotational Killing vectors the bundle still deforms. We demonstrate an example of an isotropic and exponentially expanding cross-section that finally deflates or forms a caustic.  Furthermore, using the 1+3 covariant formalism we investigate the expansion dynamics of the congruence. For certain geometric restrictions we retrieve the Raychaudhuri equation where a color-curvature coupling is revealed. The condition to prevent the focusing of neighboring particles is given and is more likely to fulfilled in highly curved regions.  Then, we introduce the Levi-Civita connection for the osculating Riemannian metric and develop a (spatially) isotropic and homogeneous dust-like model with a non-singular bounce.
\end{abstract}


\maketitle

\section{Introduction}
In Einstein's theory of gravity the 4-dimensional space-time is described by a pseudo-Riemannian manifold. Locally, the curvature effect is diminished and the Lorentz symmetry is always restored. However, when we investigate non-local phenomena, the curvature effect distorts the distance module  and we depart from the Minkowski line-element \footnote{In a particular point P(x) of a reference geodesic $\gamma(s)$ that serves as the origin of a Riemann coordinate system, the metric is given by the series approximation $g_{ab}=n_{ab}-\frac{1}{3}R_{acbd}x^cx^d+O(x^3)$.}.  The most relevant manifestation of this curvature contribution is the gravitational redshift of light. When we consider  hypothetical modifications of general relativity (GR) we can violate the Lorentz symmetry locally and/or modify the way we distort the distance module in a non-local region. These scenarios are the common playground for phenomenological models of quantum/emergent gravity.

 From a geometric perspective, we can introduce such GR modifications by dropping the restriction that the line-element should only depend on quadratic terms with respect to the coordinate increments. This non-trivial ``symmetry breaking'' is the essence of the velocity dependent geometry named as Finsler \cite{RundFB, Bao, Shen, ShenAms, Miron}. In particular, the lack of the quadratic restriction directly introduces Lorentz violations (LV) \cite{Ish,Kostelecky:2011qz,Bogoslovsky}  since the space-time locally is no longer invariant under the boosts of special relativity. Also, the global characteristics of the space-time geometry are modified as a consequence of the deformed length interval.

Extending the notion of distance in the field of non-quadratic metric functions brings into play an extra geometric property, apart from the curvature, related to departures from the quadratic measurement. This extra property of the Finsler manifold, often referred as color \cite{Shen, Lands}, directly introduces LV effects within a geometric framework. It may be considered as an effective manifestation of quantum/emergent gravitational phenomena. Moreover, the color of the manifold  enforces us to introduce two types of covariant derivatives in order to properly define the parallel displacement. The first covariant derivative (derivative of the horizontal space) generalizes the Riemannian derivative, while the second  (derivative of the vertical space) is a pure Finslerian entity. Apparently, this ``exotic'' differentiation of Finsler geometry  gives rise to three distinct Ricci identities that result from mixing up the vertical and horizontal  derivatives. This set of covariant identities encodes all the information about the kinematics of Finslerian flows.

Finsler spaces can be equivalently described as the geometry of a velocity dependent metric where the distance between two points is invariant under reparametrization of the connecting curve. In the literature velocity dependent metrics have been studied from the onset of general relativity, especially in geometric optics and analogue gravity.  One of the main ``privileges'' of Finsler space-times is that they naturally induce modified dispersion relations \cite{Girelli:2006fw}. Recently, this type of geometry appeared in different perspectives of quantum gravity (QG) (see for example \cite{Barcelo:2005fc} and references therein). For example:  in Horava-Lifshitz gravity rays may follow Finsler geodesics  \cite{Sindoni:2009vj}, within the context of the the ``stringy'' space-time foam where the D-particles recoil with the world-sheet \cite{Mavromatos:2010ar}, in Cohen and Glashow's very special relativity \cite{Cohen:2006ky}, in bi-metric theories of gravity \cite{Barcelo:2005fc, Skakala:2010hw} and in holographic fluids \cite{Leigh:2011au}. There are also studies concerning extensions of relativity theory within the framework of pseudo-Finsler structures dating back to the 40's \cite{Randers,Beem} with several issues still remain open \cite{Genfin}.

In GR the internal motion of a flow is ``encrypted'' in the failure of the deviation vector to be  parallel transported along the congruence. Following the mathematical definition of the Lie drag along the flow we can track the evolution of expansion, shear and vorticity with the aid of the Ricci identity. This elegant and straightforward derivation leads to Raychaudhuri's equation that played a keynote role in the development of GR (for some recent works see \cite{Grraych}). It shed light to crucial questions like gravitational collapse and singularities, gravitational redshift and lensing, accelerating expansion and inflation. This discussion is carried on to various branches of QG where intriguing effects are reported \cite{QGraych}. In the same line, the recent entry of Finsler geometry in emergent aspects of QG creates the need for further insight in the volume evolution of a congruence.

Motivated by the mentioned research works we will investigate the deformation of time-like congruences following E. Cartan's metrical approach \cite{RundFB, Miron, Cartan}.  The kinematics of Finsler geometry includes several complications coming from the three different Ricci identities and the anisotropic corrections in the Lie and absolute derivative. One must solve the global problem of neighboring curves and their deviation vector by imposing specific symmetries for the medium. In particular restrictive cases we demonstrate how we can retrieve a closed form expression for the expansion of neighboring particles. The generalized Raychaudhuri's equation is derived, where  a color-curvature coupling is revealed that becomes more important in highly curved regions. The simple case of a dust-like fluid moving along irrotational geodesics gives back a ``counter gravity'' condition where the focusing of matter is prevented.  Finally, the limit of osculating Riemannian metric is considered for a Randers  space-time.  In  highly spatially curved regions the Finsler contribution can dominate and lead to a bouncing  Friedmann-Robertson-Walker (FRW) model.

\section{Finsler space-times}

 Let $\sigma(\tau)$ be a smooth curve in a manifold $\mathcal{M}$ connecting two space-time points. The distance between  the points is given by the following integral over the connecting curve and its tangent direction
 \begin{equation}
 \mathfrak{L}=\int F\left(\sigma(\tau),\dot{\sigma}(\tau)\right)d\tau.
 \label{dist}
 \end{equation}
The metric function $F$ is  defined over the tangent bundle $\mathcal{TM}$. By imposing that the distance (\ref{dist}) is independent of the curves parametrization $\tau$, we get
\begin{equation}
F(x,\lambda y)=\lambda F(x,y),\;\;\;\lambda>0.
\end{equation}
where $y^a=\frac{dx^a}{d\tau}$ is the tangent vector to the $\sigma(\tau)$-curve. In other words, the metric function $F(x,y)$ is homogeneous of degree one with respect to the coordinate increments $y^a$. Using Euler's theorem of homogenous functions we retrieve the formula
\begin{equation}
F^2(x,y)=g_{ab}(x,y)y^ay^b,
\label{finmet}
\end{equation}
where $g_{ab}$ is the metric tensor and is given by the relation
\begin{equation}
g_{ab}(x,y)=\frac{1}{2}\frac{\partial^2F^2}{\partial y^a \partial y^b}.
\label{fmeten}
\end{equation}
 Furthermore, a Riemann manifold is retrieved if $F(x,y)$ is quadratic with respect to $y^a$. Then, the double partial derivative of the previous relation removes the $y^a$-dependence of the metric. In that sense, Finsler geometry generalize the Riemannian case within the field of non-quadratic metric functions. The Euler-Lagrange equations for the extremal of $F(x,y)$  give back the usual relation
 \begin{equation}
 \frac{dy^a}{d\tau}+2G^a=0,
 \label{eula}
 \end{equation}
  where the second term in the lhs is usually called the spray induced by $F(x,y)$ and is related to the Christoffel symbols by the relation
  \begin{equation}
  G^a=\frac{1}{2}\gamma^a_{\;\;bc}y^by^c.
  \label{spray}
  \end{equation}
  The extremal curves (\ref{eula}) define a space of paths for the Finsler manifold.

   The essential difference in the Finslerian approach is that the quadratic restriction on the metric function is dropped. Consequently, the metric tensor depends on the variable $y^a$ and each tangent space $\mathcal{T}_x\mathcal{M}$ is no-longer equipped with ellipsoidal unit balls. Instead, there will be a locus given by the restriction $F(x,y)=1$ (see rel.(\ref{finmet})). This non-ellipticity of the unit balls gives an extra property to the Finslerian manifold, apart from curvature, referred as color. The main quantity that measures the color of our structure is the Cartan torsion tensor
 \begin{equation}
 C_{abc}=\frac{1}{2}\frac{\partial g_{ab}}{\partial y^c},
 \end{equation}
 since the condition $C_{abc}=0$ implies a Riemann space. In that sense, a Finsler space is a colorful curved manifold, while a Riemann space is curved but entirely white  \footnote{The counter example of  curved and white (Riemann) geometry is the flat and colored case, $F=F(y)$. In fact, this general class of norms describes the geometry of each tangent space $\mathcal{T}_x\mathcal{M}$ of the Finsler manifold. They are considered as standard model extensions since particles moving in the colored set up, $F^2=g_{ab}(y)y^ay^b$, will obey modified mass-shell conditions. }. Recently, this extra property of the $\mathcal{F}_n$ manifold  drew some attention in the community of theoretical physics since it is related to Lorentz violating scenarios (see for example \cite{Kostelecky:2011qz}, \cite{Bogoslovsky} and \cite{Girelli:2006fw}).

  We consider the position space together with the $y^a$ increments, that solve relation (\ref{eula}), as the fundamental variables of the physical space-time.
  In other words we replace the Riemann space-time with a Finsler one where the position space gives place to the element of support $(x,y)$ \cite{RundFB},\cite{Cartan}. Thus, tensor fields will depend on the coordinates of the tangent bundle $\mathcal{TM}$ induced by the local chart of the base manifold $\mathcal{M}$. In the spirit of \cite{Beem}, the square magnitude of a first rank tensor $u^a$ in $x$ is given by $g_{ab}(x,y)u^au^b$ and it separates to three classes, spacelike, null and timelike for negative, zero and positive values respectively.

 The tangent bundle $\mathcal{TM}$ has a local coordinate base $\{\frac{\partial}{\partial x^a},\frac{\partial}{\partial y^a}\}$. However, under a transformation on $\mathcal{TM}$ induced by a coordinate change on $\mathcal{M}$, the elements ${\partial\over \partial x_a}$ do not transform linearly. This problem is solved by using instead the {\it adapted frame}  $\{\frac{\delta}{\delta x^a},\frac{\partial}{\partial y^a}\}$, with
 \begin{equation}
 \frac{\delta}{\delta x^a}=\frac{\partial}{\partial x^a}-N_a^{\;\;b}\frac{\partial}{\partial y^a},
 \label{nonlin}
 \end{equation}
 where $N_a^{\;\;b}={\partial G^a\over\partial y^b}$ is the non-linear connection. Under the same argument, the adapted frame in the cotangent bundle $\mathcal{T^{*}M}$ is $\{dx^a,\delta y^a\}$, with
 \begin{equation}
 \delta y^a=dy^a+N_a^{\;\;b}dx_a.
 \end{equation}
 The presence of the non-linear connection imposes that the adapted frame is non-holonomic in the sense that the following commutation formulas hold
 \begin{eqnarray}
 &&[{\delta\over\delta x^a},{\delta\over\delta x^b}]=R^c_{\;\;ab}{\partial\over\partial y^c},\\
 &&[{\delta\over\delta x^a},{\partial\over\partial y^b}]=G^c_{\;\;ab}{\partial\over\partial x^c},\\
 &&[{\partial\over\partial y^a},{\partial\over\partial y^b}]=0,
 \end{eqnarray}
 where we used the definitions
 \begin{equation}
 R^a_{\;\;bc}={\delta N^a_{\;\;b}\over\delta x^c}- {\delta N^a_{\;\;c}\over\delta x^b},
 \label{rtor}
 \end{equation}
 and
\begin{equation}
G^a_{\;\;bc}=\frac{\partial N^a_{\;\;b} }{\partial y^c}.
\label{Gtor}
\end{equation}
Note, that the tensor field (\ref{rtor})  is also present in the Riemann limit  and reflects the gravitational field. On the other hand, the $G^a_{\;\;bc}$ components  monitor the ``mess up'' between the coordinates of the base manifold $\mathcal{M}$ and the elements of the tangent bundle $y^a$.

\section{ Aspects of Finslerian congruences}
Consider a family of curves $\gamma(s)$ and their tangent vector field $u^a$ in a manifold $\mathcal{M}$. The parallel displacement of a vector $X^a$ along the $u^a$-bundle is given by the following formula
\begin{equation}
\frac{DX^a}{ds}=\frac{dX^a}{ds}+P^a.
\label{pard}
\end{equation}
The components $P^a$ are to be specified by the geometric properties of $\mathcal{M}$. We consider the mathematical definition of the deviation vector $\xi^a$  as the tangent of a second flow given by the Lie drag of a connecting curve along  the congruence. The failure to parallel transport the deviation vector along the flow monitors the internal motion of the $u^a$-family, namely
\begin{equation}
\frac{D\xi^a}{ds}=B^a_{\:\: b}\xi^b+A^a_{bc}\xi^b\xi^c+...
\label{dv1}
\end{equation}
The tensor fields in the rhs of the above can take explicit expressions depending on the geometric assumptions we impose. The common example is the Riemann case given in the next subsection.

\subsection{White curved space-times}\label{riem}
When the manifold is Riemannian the parallel displacement (\ref{pard}) involves only corrections given by the linear connection. Taking into account that the deviation vector is Lie transported along the flow, $\pounds_u \xi^a=0$, the absolute derivative of $\xi^a$ can be treated as
\begin{equation}
\dot \xi^a\equiv\frac{D\xi^a}{ds}=u^b\nabla_b\xi^a=\xi^b\nabla_bu^a.
\end{equation}
Combining the previous expression with relation (\ref{dv1}) we find that only first order terms are involved in our approximation. Using the 1+3 covariant formalism \footnote{Given a time-like vector, $u^au_a=1$, we can define the tensor $h_{ab}=g_{ab}-u_au_b$ that projects orthogonal to $u_a$. With the later we can decompose a vector to its irreducible parts $X_a=Xu_a+\mathcal{X}_a$. The scalar factor corresponds to the components along the flow (time-like part) $X=X_au^a$, while the vector part $\mathcal{X}_a=h_a^{\;\;b}X_b$ is the projection on the instantaneous rest space of the fundamental observer (space-like part).}, we can write into its irreducible parts the second rank tensor that encodes all the relevant information
\begin{equation}
B_{ab}=\nabla_bu_a=\frac{1}{3}\Theta h_{ab}+\sigma_{ab}+\omega_{ab}+A_au_b,
\label{RBab}
\end{equation}
where we use the definitions with respect to the 3-D covariant operator ${\rm D}_a=h_a^{\;\;b}\nabla_b$, for the expansion of a volume element $\Theta={\rm D}^au_a$, for the shear $\sigma_{ab}=2{\rm D}_{ (b}u_{a) }-{1\over3}\Theta h_{ab}$, for the vorticity $\omega_{ab}={\rm D}_{[ b}u_{a]}$ and the four acceleration $A_a=\dot u_a$  \cite{Relcos}.

In Riemann geometry the covariant derivative on a vector field fails to commute due to the curvature of space-time. The tensorial Ricci identity reads
\begin{equation}
2\nabla_{[a}\nabla_{b]}u_c=R_{abcd}u^d,
\label{RRicci}
\end{equation}
and decomposes to three propagation and three constrain equations \cite{Relcos} for the irreducible parts of relation (\ref{RBab}). The most relevant equation of this set is Raychaudhuri's equation that monitors the evolution of the average volume of a spatial 3-D element $d\mathcal{V}$. It results directly from the Ricci identity (\ref{RRicci}) by taking the trace and projecting along the $u^a$-family
\begin{equation}
\dot{\Theta}+\frac{1}{3}\Theta^2=-R_{ab}u^au^b-2(\sigma^2-\omega^2)+{\rm D}_aA^a-A_aA^a,
\label{RRay}
\end{equation}
where $\sigma^2=\sigma^{ab}\sigma_{ab}/2$ and $\omega^2=\omega^{ab}\omega_{ab}/2$ and the last two terms are representatives of external forces. Given an initially expanding phase of the flow ($\Theta>0$) we directly conclude that positive terms in the rhs will accelerate the expansion and negative ones will decelerate.  The statement is reversed in case of a contracting phase. Notice, that the evolution law (\ref{RRay}) is a pure geometric relation. It is valid for any theory of gravity that assumes a Riemann manifold for the position space. We can indirectly connect the evolution formula (\ref{RRay}) with  the energy and momentum of matter through the field equations and the conservation relations. In the context of GR Raychaudhuri's equation reflects the attracting nature of gravity, since for ordinary matter the curvature contribution always assists the contraction  \cite{Grraych}.

\subsection{Parallel displacement and Ricci identities }
As we discussed, in Finsler spaces tensor fields no longer depend only on the position but instead they depend on the {\it element of support} $(x,y)$ \cite{RundFB,Cartan}. Therefore, the properties  of the tangent bundle $\mathcal{TM}$ is of central importance. The absolute differentiation of a vector field $X^a$ will contain extra contribution along the tangent displacement $dy^a$. In a local expression we can write down
\begin {equation}
DX^a=dX^a+\Gamma^{a}_{\;\;bc}X^bdx^c+C^{a}_{\;\;bc}X^bdy^c,
\label{absder}
\end{equation}
where the coefficients $C^{a}_{\;\;bc}$ and $\Gamma^{a}_{\;\;bc}$ are functions of the element of support. We will restrict our analysis only along the normalized time-like direction $u^a={dx^a\over ds}$. This congruence plays an analogous role with the fundamental observer of the 1+3 covariant formalism.

By imposing the  metricity condition $Dg_{ab}=0$, along any direction $u^a$ we can recast relation (\ref{absder}) in a covariant form by using the horizontal and vertical split
\begin{equation}
\dot X^a\equiv\frac{DX^a}{ds}=X^a_{\;\;| b}u^b+X^a|_{ b}F\frac{Dl^b}{ds},
\label{absD}
\end{equation}
where
\begin{eqnarray}
&&X^a_{\; \; |b}=  \frac{\delta X^a}{\delta x^b}+L^a_{\;\; bc}X^c\label{Dhor},\\
&&X^a|_{ b}=  \frac{\partial X^a}{\partial y^b}+C^a_{\;\; bc}X^c \label{Dvert},
\end{eqnarray}
and $l^a=y^a/F$ is the unit vector in the direction of the element of support $(x,y)$ while $Dl^a=dl^a+L^a_{\;\;bc}l^bdx^c$. Note, that for $X^a=\kappa l^a$ relation (\ref{absD}) implies $l^a_{\:\:|b}=0$ and the spray coefficients (\ref{spray}) determine the integral curves. We will refer to the operator (\ref{Dhor}) as the horizontal covariant derivative and to (\ref{Dvert}) as the vertical one. Also, the connection coefficients in (\ref{Dvert}) satisfy the following relation
\begin{equation}
L^a_{\; \; bc}=\frac{1}{2}g^{ad}\left( \frac{\delta g_{db} }{\delta x^c}+\frac{ \delta g_{dc} }{\delta x^b}-\frac{ \delta g_{bc} }{ \delta x^d } \right).
\end{equation}
In that case a parallel displaced vector field $DX^a=0$ keeps its length invariant. The covariant expression (\ref{absD}) is the evolution law for any first rank tensor. The  vertical part (\ref{Dvert})  is a direct result of the distorted ellipsoid condition (\ref{finmet}). Together with the non-linear connection (\ref{nonlin}) monitors the LV effect on the evolution of the $X^a$-flow.

The splitting of the propagation relation (\ref{absD}) allows us to formulate a curvature theory in the footsteps of Riemann geometry. Using the vertical and horizontal covariant derivatives we retrieve the following generalized Ricci identities
\begin{eqnarray}\label{Ric1}
&X^a_{\;\;| b| c}-X^a_{\;\;| c| b}&=  X^dR^{\;\; a}_{d\;\;bc}-X^a|_dR^d_{\;\;bc}, \\\label{Ric2}
&X^a|_b|{}_c-X^a|_c|_b&=  X^dS^{\;\; a}_{d\;\;bc},
\end{eqnarray}
and
\begin{equation}
X^a_{\;\;{\mid} b}|{}_c-X^a|_{c}{}_{| b}=X^dP^{\;\; a}_{d\;\;bc}-X^a_{\;\;|d}C^d_{\;\;bc}
-X^a|_dP^d_{\;\;bc},
\label{Ric3}
\end{equation}
where the curvature components are defined by
\begin{eqnarray}
&&R_{d\;\;bc}^{\;\;a}=\frac{\delta L^a_{\;\;db}}{\delta x^c}-\frac{\delta L^a_{\;\;dc}}{\delta x^b}+L^e_{\;\;db}L^a_{\;\;ec}-L^e_{\;\;dc}L^a_{\;\;eb}+C^a_{\;\;de}R^e_{\;\;bc},\\
&&S_{d\;\;bc}^{\;\;a}=\frac{\partial C^a_{\;\;db}}{\partial y^c}-\frac{\partial C^a_{\;\;dc}}{\partial y^b}+C^e_{\;\;db}C^a_{\;\;ec}-C^e_{\;\;dc}C^a_{\;\;eb},\\
&&P_{d\;\;bc}^{\;\;a}=\frac{\partial L^a_{\;\;db}}{\partial y^c}-C^a_{\;\;dc|b}+C^a_{\;\;de}P^e_{\;\;bc}
\end{eqnarray}
and the torsion-like tensor field $P^a_{\;\;bc}$ is given by the following expression
\begin{equation}
P^a_{\;\;bc}=\frac{\partial N^a_{\;\;b} }{\partial y^c}-L^a_{\;\;cb}.
\end{equation}
The Ricci identities (\ref{Ric1})-(\ref{Ric2}) enclose the relevant information about the non-linear kinematics of the integral curves since the torsion-like tensors $C^a_{\;\;bc},R^a_{\;\;bc}$ and $P^a_{\;\;bc}$ are fully determined by the fundamental function $F(x,y)$.

The irreducible decomposition of relations (\ref{Ric1})-(\ref{Ric3}) will give back a set of covariant relations \cite{Stavrin}. However, this is not enough to determine the actual deformation of the $X^a$-curves. One must further proceed to connect the aforementioned relations with the actual internal motion of the congruence. In the general case, it seems difficult to retrieve closed form expressions for the evolution of expansion, shear and vorticity. Therefore, to get the deformation of a cross-section we need to solve the global problem and track back the kinematics of the medium, either by considerable simplifications or by using numerical techniques.

To further dig  into this problem, consider a family of time-like curves and their tangent vector $u^a$ in a Finsler space-time. By virtue of  the absolute differentiation (\ref{absD})  the propagation equation for the deviation vector is written in the form
\begin{equation}
\dot{\xi^a}=\xi^a_{\;\;| b}u^b+\xi^a|_{ b}F\frac{Dl^b}{ds}
\label{absDxi}
\end{equation}
 Notice, the extra contribution in the evolution of the deviation vector, coming from the vertical operator along the $\frac{Dl^a}{ds}$-direction. The component of $\frac{Dl^a}{ds}$ along $\xi^a$ will involve up to second order terms with respect to $\xi^a$ in the kinematics of the medium (\ref{dv1}) (for a similar discussion on higher order terms in the Riemannian framework see \cite{schutz}). The most relevant contribution of the second order terms is expected when $\frac{Dl^a}{ds}$ is parallel to the deviation vector. On the other hand the second order terms will fade away if $\frac{Dl^a}{ds}$  is parallel transported along the congruence.  Using the mathematical definition of the deviation vector as the Lie transported field along the $u^a$-congruence we shall retrieve further analytic expressions for the deformation of the integral curves in these two limiting cases of $\frac{Dl^a}{ds}$ (subsections (\ref{sphsym}) and (\ref{Raysub}) respectively).

\section{Lie derivative and Killing vectors}
 The most closely related  to the Riemannian case definition  of the Lie derivative results from  the infinitesimal transformation on the position coordinates of $\mathcal{M}$
 \begin{equation}
 \bar{x}^a=x^a+v^a(x)d\tau,\label{infx}
 \end{equation}
 where $d\tau$ is an infinitesimal constant and $v^a(x)$ is a vector field defined over a region of the base manifold \cite{RundFB}. The above displacement assigns at each point $x^a$ a shift on the position space $dx^a=v^a(x)d\tau$ that implies a corresponding variation of the $y^a$ components of the element of support
 \begin{equation}
 \bar{y}^a=y^a+\left(\frac{\partial v^a}{\partial x^b}y^b\right)d\tau.\label{infy}
 \end{equation}
 Following the usual procedure \footnote{The infinitesimal displacement of $X^a(x,y)$ for the variations (\ref{infx}) and (\ref{infy}) gives back the relation $dX^a=\frac{\partial X^a}{\partial x^b}v^bd\tau+\frac{\partial X^a}{\partial y^b}\left(\frac{\partial v^b}{\partial x^c}y^c\right)d\tau$ and combined with the variation of the vector with respect to the transformation law $\bar{X}^a=\frac{\partial \bar{x}^a}{\partial x^b}X^b$ gives back the formula for the Lie derivative.  } for the Lie derivative of a vector field $X^a(x,y)$ we lead to the formula
\begin{equation}
\pounds_v X^a=X^a_{\;\;|b}v^b-v^a_{\;\; |b}X^b+\frac{\partial X ^a}{\partial y^b}(v^b_{\;\; |c}y^c).
\label{Lieder}
\end{equation}
The third term on the rhs of the previous relation  comes from the preferred direction imposed by the element of support. However, the reader should keep in mind that relation (\ref{Lieder}) is a special case, since the Lie derivative is defined along the vector $v^a(x)$ that is independent of the coordinate increments $y^a$.  The condition $\pounds_{v}g_{ab}=0$ implies the equation for the Killing vectors
\begin{equation}
v_{a|b}+v_{b|a}+2FC_{abc}v^c_{\;\;|d}\,l^d=0.
\label{Killng}
\end{equation}
Thus, the Killing solutions are kinematically enriched compared to  general relativity since Cartan's torsion tensor $C_{abc}$ and the element of support  implies expansion, shear and vorticity for the bundle.

 Consider the 4-velocity of a time-like congruence $u^a$ together with the deviation vector $\xi^a$ and let us assume that they both depend only on the position  $x^a$. In that case, relation (\ref{Lieder}) takes the simplified form
 \begin{equation}
 \pounds_u\xi=\xi^a_{\;\;|b}u^b-u^a_{\;\; |b}\xi^b=0.
 \label{Lzr}
 \end{equation}
 For a parallel displaced element of support, $\frac{Dl^a}{ds}=0$, the vertical contribution disappears from relation (\ref{absDxi}) and the horizontal derivative plays the same role with the Riemann covariant derivative. Therefore, the internal deformation of the $u^a$-flow is fully described by the irreducible parts of $B_{ab}=u_{a|b}$.  Thus, assuming a geodesic time-like congruence and using (\ref{absDxi}) together with (\ref{Killng}) and (\ref{Lzr}) we get
\begin{equation}
 \Theta_{ab}+F\mathcal{C}_{abc}\,(\sigma^c_{\;\;d}+\omega^c_{\;\;d})\,\ell^d=0,
 \end{equation}
where  we define $\mathcal{C}_{abc}=h_a^{\;\;d}h_b^{\;\;e}h_c^{\;\;f}C_{def}$ and $\Theta_{ab}=\frac{1}{3}\Theta h_{ab}+\sigma_{ab}$ in analogy to the Riemann case for the horizontal covariant derivative, while $\ell^a=h^a_{\:\: b}l^b$ is the space-like part of the unit vector $l^a$. Thus, for a time-like congruence $u^a(x)$  that is a Killing vector and the deviation vector depends solely in the position, the expansion and shear are non-zero if the element of support points to a preferred direction in space. Note, that shear free and irrotational Killing vectors of the previous example are always non-expanding.

\section{Finslerian curves and  volume evolution    }
The only geometric tool in hand to measure the relative motion of a family of curves $\gamma (s)$ is the parallel displacement of the deviation vector. As we already mentioned this displacement is directly related to the covariant derivative of the tangent vector $u^a$. However, in Finsler geometry  the covariant derivative is replaced by the horizontal and vertical decomposition of the absolute derivative. This is a natural result, since the abandonment of the quadratic restriction forces us to replace the  position space $x^a$ with the element of support $(x^a,y^a)$. Therefore, to isolate the volume evolution of a Finslerian congruence we need to take into account all the curvature  together with the torsion tensors (\ref{Ric1})-(\ref{Ric3}) and up to second order terms in relation (\ref{dv1}). The analytical treatment of this problem seems a very complicated task but we can get some insight through some geometric assumptions.

Firstly, let us prove that second order terms in relation (\ref{dv1}) enter the kinematics. Consider a time-like flow $u^a$ and its deviation vector $\xi^a$ that depends  only on $x^a$.  The parallel displacement of the unit direction $l^a$ along $u^a$ decomposes to a component along and perpendicular to the deviation vector $\xi^a$
\begin{equation}
\frac{Dl^a}{ds}=\mathfrak{\Lambda} \xi^a+\mathfrak{\Lambda}^a.
\label{Dl1p3}
\end{equation}
To focus on the second order terms we assume that the transverse part of (\ref{Dl1p3}) vanishes,  $\mathfrak{\Lambda}^a= 0$. Note, that this assumption implies $l^a\xi_a=0$ since $l^a=y^a/F$ by construction. Then, by virtue of (\ref{Dhor}),(\ref{absDxi}) and (\ref{Lzr}) we get the following relation
\begin{equation}
\frac{D\xi^a}{ds}=u^a_{\;\;|b}\xi^b+\mathfrak{\Lambda}FC^a_{\;\;bc}\,\xi^b\xi^c,
\label{DxiTor}
\end{equation}
which proves our previous statement that second order terms are naturally involved in the evolution of the deviation vector. It is evident from relation (\ref{DxiTor}) that an analytic expression for the expansion, shear and vorticity of the congruence is unapproachable.

The second order terms of $\xi_a$ complicate the picture and even in simple geometric configurations non-linear effects will emerge. Even if we consider the isotropic condition for the first order terms $u_{a|b}=Bh_{ab}$, Cartan's tensor will change the initial direction of $\xi^a$. Therefore, an isotropic cross-section will eventually be distorted.  In the following two subsection, we study two particular covariant examples of relation (\ref{DxiTor}). In the first one, we impose spherical symmetry in the evolution law (\ref{DxiTor}) and retrieve particular solutions with respect to the affine parameter of the congruence. In the second one, we restrict our analysis to a particular class of curves where the anisotropic unit direction $l^a$ is parallel displaced and recover an evolution formula for the expansion of the bundle.

 \subsection{Spherical symmetric evolution}\label{sphsym}
 Assume that  Cartan's torsion tensor takes its reducible form \footnote{ Cartan's torsion tenor is always reducible in case of a Randers or Kropina space. It is also reducible for any Finsler metric in case of a surface.}
\begin{equation}
C_{abc}=C_{(a}\mathfrak{h}_{bc)},
\label{redca}
\end{equation}
where $C^a=C^a_{\;\;bc }g^{bc}$ is often referred as Cartan's torsion vector. Also,  $\mathfrak{h}_{ab}=g_{ab}+cl_al_b$ is the \emph{angular tensor} that projects orthogonal to $y_a$ and $c=\mp1$ for a space-like or time-like unit direction respectively. Apart from the anisotropic contribution from Cartan's tensor in relation (\ref{DxiTor}) there is an isotropic effect along the direction of $\xi^a$. In particular, we recover an isotropic evolution for the cross-section  when $C_a$ is parallel to the deviation vector $C_a=\tilde{C}\xi_a$ and the simple relation $u_{a|b}=Bh_{ab}$ holds. Then, by virtue of these assumptions and using relations (\ref{DxiTor}) and (\ref{redca}) we arrive at the expression
\begin{equation}
\frac{D\xi^a}{ds}=(B+ C\xi^2)\xi^a,
\label{NODExi}
\end{equation}
  with $C=\mathfrak{\Lambda}F\tilde{C}$. Thus, in the above example the deviation vector  always points at the same direction as it is parallel displaced with respect to the $\gamma(s)$ curves.   An initially spherical symmetric region of space will sustain its shape as it moves along the $u^a$-congruence. In that case, the magnitude of the deviation vector is always the radius $r$ of an isotropic 3D-section. Contracting relation (\ref{NODExi}) along $\xi^a$ we derive the  evolution law for the radius
\begin{equation}
\dot{r}=(B+Cr^2)r,
\label{Vode}
\end{equation}
where the dot operator denotes the absolute derivative with respect to $s$. The second term in the rhs of relation (\ref{Vode}) involves non-linear corrections to the expansion dynamics through Cartan's torsion vector. When Cartan's tensor vanishes, we retrieve the vorticity and shear free expansion of the Riemannian limit where $B=\frac{1}{3}\Theta$. In that case a constant expansion rate guarantees an exponential solution for the volume element. The exponential behavior is closely related to inflationary scenarios of the early universe.

\begin{figure}
\begin{center}
\includegraphics[width=0.48 \textwidth, angle=0]{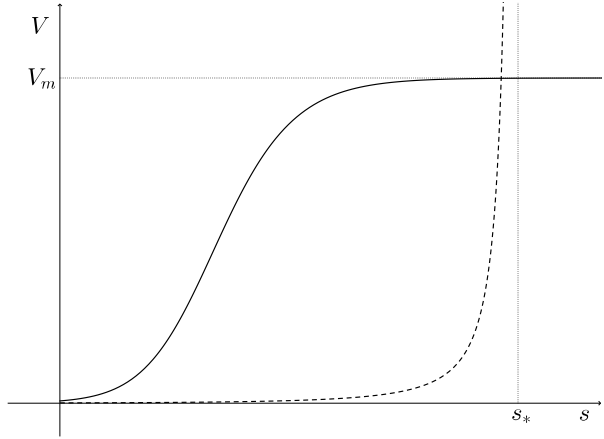}
\end{center}
\caption{\small  Free scale volume evolution  for an initially expanding cross-section with respect to the parameter  of the fundamental $u^a$-congruence. The solid line corresponds to  negative values of $C$. An initially exponential expansion gradually turns to a decelerating phase that finally ``freeze'' to a constant value $V_{m}=(-\frac{B}{C})^{3/2}$. On the other hand, positive values of $C$ (dashed line) set an upper value $s_*$ for the arc-length where the volume element reach the caustic.      }\label{volfig}
\end{figure}

To demonstrate how a simple inflationary model can be affected by the non-linear Finsler contribution let $B=const$ and $C=const$.  This limit retrieves the aforementioned behavior for $C\rightarrow0$. The solution of relation (\ref{Vode}) leads to the formula for the volume of an isotropic cross-section of the $u^a$-congruence
\begin{equation}
V=\left(\frac{B}{c_1Be^{-Bs}-C}\right)^{3/2},
\label{vol}
\end{equation}
where $c_1$ is a constant of integration. In the Riemannian approximation, the  case of $B<0$ corresponds to an exponential contraction which tends to the singular value, $V\rightarrow0$. For the Finslerian volume element (\ref{vol}) given the conditions $B<0$ and $C>0$ we get an additional phase of decelerating contraction for small values of the parameter $s$. On the other hand, for $B<0$ and $C<0$ the element monotonically contracts. In both cases of contraction the singular fate, $V\rightarrow0$, is always unavoidable. The inflationary behavior in general relativity is retrieved  for $B>0$ where a 3-sphere eternally expands exponentially. In relation (\ref{vol}), for $B>0$ and $C<0$  an initially inflationary expansion turns to a decelerating one that leads to a "frozen" state, $V_m=(-\frac{B}{C})^{3/2}$. Furthermore, for $B>0$ and $C>0$  and restricting the values of $V$ to be real, the parameter $s$ is bounded by an upper value $s_*=\frac{1}{B}\ln\frac{c_1B}{C}\,$. The volume element tends to a caustic as we approach $s_*$ since its radius reach infinity within a finite time interval. The two cases of expansion are depicted in figure-\ref{volfig}.

\subsection{Raychaudhuri's equation}\label{Raysub}
The intrinsic non-linear nature of the Finsler manifold  disapproves the extraction  of the propagation equations for the kinematical properties of the $u^a$-bundle. In the general case the evolution of the expansion cannot be expressed in  closed form. This is evident even in the simplified example of the previous section (see relation (\ref{Vode})). The main suspect for this complication is the vertical part of the absolute derivative. In order to derive  Raychaudhuri's equation we are forced to impose some geometric constraints. The first drastic assumption taken into account concerns the absolute derivative  of the unit direction $l^a$.

 Precisely, let us consider a family of time-like flow lines along which $l^a$ is parallel displaced, $\frac{Dl^a}{ds}=0$. In that case, relation (\ref{absDxi}) gives back the evolution of the deviation vector, $\dot\xi_a=\xi_{a|b}u^b$.  The unit element is written into its irreducible parts as
\begin{equation}
l_a=lu_a+\ell_a,
\end{equation}
and since the absolute derivative along $u^a$ vanishes we retrieve the expression for the 4-acceleration of the congruence
\begin{equation}
lA_a=-\dot{\ell}_{\langle a \rangle}.
\label{dla}
\end{equation}
 Using relation (\ref{dla}) together with the limit that the connecting vector $\xi^a$ depends only on $x$, the Lie derivative (\ref{Lzr}) implies that only first order terms are involved in the rhs of (\ref{absDxi}). Moreover, using relations (\ref{absDxi}) and (\ref{Lzr}) for a class of curves where the anisotropic unit direction $l^a$ is parallel displaced ($\dot{l^a}=0$), we retrieve the relation
\begin{equation}
\dot{\xi}^a=u_{a|b}\xi^b.
\end{equation}
 Hence, in our limiting case we can track the internal motion of the medium with the aid of the horizontal derivative
\begin{equation}
B_{ab}=u_{a|b}.
\end{equation}
in analogy to the Riemannian case discussed in subsection (\ref{riem}).
The irreducible parts of the above expression monitor the expansion, shear and vorticity of the congruence. Their definitions are similar to the Riemannian case (\ref{RBab}) although the Levi-Civita connection $\nabla_a$ is replaced by the horizontal operator that involves the non-linear connection.

Furthermore, the time-like part of the contracted Ricci identity gives back the propagation equation for the expansion
\begin{eqnarray}\nonumber
\dot{\Theta}+\frac{1}{3}\Theta^2&=&-R_{ab}u^au^b-2(\sigma^2-\omega^2)-\mathfrak{T}_{ab}u^au^b\\
&&+\frac{1}{l^2}\dot{\ell}^a({\rm D}_al-\dot{\ell}_{\langle a\rangle})-\frac{1}{l}{\rm D}^a\dot{\ell}_a,
\label{Raych1}
\end{eqnarray}
 where  $\sigma^2=\sigma^{ab}\sigma_{ab}/2$ and $\omega^2=\omega^{ab}\omega_{ab}/2$ are respectively the scalar square magnitudes of the shear and vorticity, while $\mathfrak{T}_{ab}=C^{\;\;\;\;d}_{ac}l^eR^{\;\;c}_{e\;\;db}$ represents a color-curvature coupling coming from the anholonomy of the basis and the intrinsic anisotropy of the metric. The last three terms are due to the acceleration given in relation (\ref{dla}) and correspond to non-geodesic motion. In the rhs of relation (\ref{Raych1}) positive terms resist the contraction of the time-like bundle while negative terms assist the focusing of the congruence. On the other hand, in an expanding phase positive terms accelerate the deviation of neighboring flow-lines while negative terms contribute with a decelerating effect.

\section{Collapse of a dust-like fluid}
In general relativity Raychaudhuri's equation is a necessary tool to study the behavior of a self-gravitating fluid. It has a purely geometrical origin although it is indirectly correlated to the energy and momentum of matter through Einstein's field equations. When we consider irrotational ($\omega_{ab}=0$) perfect fluid configurations of ordinary matter the congruence always forms  a caustic. On the other hand, if we replace the Riemann manifold with a Finslerian we are unable to derive in the general case a closed form expression for the expansion. Also, the relation between matter and geometry remains in the sphere of speculation but with remarkable progress been made \cite{Miron,Pfeif}.

The derivation of relation (\ref{Raych1}) requires  a parallel displaced  element of support $\frac{Dl^a}{ds}=0$ along the $u^a$-bundle. Under this assumption the relevant Finsler entities that enter  Raychaudhuri's formula are the Riemann curvature (\ref{Ric1}) together with the two torsions ($C^{a}_{\;\; bc}, R^a_{\;\;bc}$) and the 4-acceleration. Furthermore, non-geodesic motion is necessarily related with the unit preferred direction through (\ref{dla}). Since, for the parallel displacement (\ref{absD}) the only operator   is the horizontal differentiation (\ref{Dhor}), it is identified  as the covariant derivative in analogy to the Riemannian case. If we impose energy and momentum conservation for our setup, then we get
\begin{equation}
T^{ab}_{\;\;\;\;|b}=0.
\label{coservT}
\end{equation}
 with $T^{ab}$ representing an imperfect fluid. One can search in the  identities of Finsler geometry for a second rank tensor that satisfies relation  (\ref{coservT}).

 Keeping close to GR, the Bianchi identities for the Riemann curvature (\ref{Ric1}) with respect to our covariant differentiation are
 \begin{equation}
 \begin{array}{ll}
 R^{\;\;b}_{a\;\;cd|e}+R^{\;\;b}_{a\;\;ec|d}+R^{\;\;b}_{a\;\;de|c}+ & \\
 +R^f_{\;\;de}P^{\;\;b}_{a\;\;cf}+R^{f}_{\;\;cd}P^{\;\;b}_{a\;\;ef}+R^{f}_{\;\;ec}P^{\;\; b}_{a\;\;df}=& 0.
 \end{array}
 \end{equation}
The nearest curvature theory to general relativity is achieved if the space-time has the property $P_{abcd}=0$. Then, after some short calculations we arrive at the following conservation law
\begin{equation}
\left(R^{(ab)}+R^{[ab]}-\frac{1}{2}Rg^{ab}\right)_{|b}=0,
\label{biatwo}
\end{equation}
where $R_{ab}=g^{cd}R_{acbd}$ and $R=g^{ab}R_{ab}$ is the Ricci tensor and scalar respectively. Notice, that due to the torsion contribution in the curvature (\ref{Ric1}) the Ricci tensor decomposes into a symmetric and antisymmetric part. Furthermore, the $P_{abcd}=0$ restriction gives back the constrain \cite{Miron} for the S-curvature
\begin{equation}
S_{abcd|e}=0\label{sconst}.
\end{equation}
As it is naturally expected the full set of the Bianchi identities for the Finslerian set up involves relations where the vertical derivative appears.  For the full set up of the structure equations the reader is referred to standard textbooks of Finsler geometry e.g. \cite{RundFB}, \cite{Bao} and \cite{Miron}. The aim of our analysis is to extract some qualitative results about the convergence of neighboring particles falling along our restricted bundle. Then,  only the horizontal derivative is involved in the expansion dynamics and  the Riemann curvature is the only curvature that directly contributes to Raychaudhuri's equation (\ref{Raych1}).

 The most trivial choice is to assume that the symmetric and antisymmetric parts of the twice contracted Bianchi identities (\ref{biatwo})  are explicitly conserved. Consequently, relations (\ref{coservT}) and (\ref{biatwo}) lead to the field equations
 \begin{equation}
 R_{(ab)}-\frac{1}{2}Rg_{ab}=\kappa T_{ab},
 \label{ein}
 \end{equation}
 where $\kappa$ is the gravitational coupling  constant. However, there are the additional constrains for the antisymmetric part of the conservation equation (\ref{biatwo}),
 \begin{equation}
 R^{[ab]}_{\;\;\;\;\;|b}=0
 \end{equation}
 and for the S-curvature given by (\ref{sconst}).
 The above field equations represent a subset of the general Finsler problem \cite{Miron,Genfin,Pfeif}. Yet,  they are compatible with the geometric assumptions we made to derive the evolution equation for the expansion (\ref{Raych1}). Relative to the fundamental observer the energy-momentum tensor of a perfect dust-like fluid is
\begin{equation}
T_{ab}=\rho u_au_b,
\label{em_rho}
\end{equation}
with $\rho=T_{ab}u^au^b$ standing for the energy density of matter. Notice, that relation (\ref{ein}) implies that the energy density will depend on the element of support.

The conservation law (\ref{coservT}) forces the matter distribution (\ref{em_rho}) to move along geodesics and the 4-acceleration vector (\ref{dla}) vanish. Then, by virtue of relations (\ref{ein}), (\ref{em_rho}) and assuming that matter falls along irrotational geodesics, we recast Raychaudhuri's equation (\ref{Raych1}) in the following form
\begin{equation}
\dot{\Theta}+\frac{1}{3}\Theta^2=-\frac{1}{2}\kappa \rho-2\sigma^2-\mathfrak{T}_{ab}u^au^b.
\label{simpRch}
\end{equation}
 The sign  in the rhs of (\ref{simpRch}) determines the condition for the formation of a caustic. In particular, the focusing of the world-lines is prevented only if the rhs of relation (\ref{simpRch}) is  dominated by positive terms, namely
\begin{equation}
\frac{1}{2}\kappa \rho+2\sigma^2<-\mathfrak{T}_{ab}u^au^b.
\label{caupr}
\end{equation}
 The term that can resist the gravitational pull of a geodesic congruence comes from the coupling between the Cartan torsion tensor and curvature. It fulfils the requirements only  if  $\,\mathfrak{T}_{ab}u^au^b<0$.   On the other hand, the color-curvature  effect adds to the self attraction of the medium when $\mathfrak{T}_{ab}u^au^b>0$. The more we deviate from the quadratic restriction the more important becomes the Cartan torsion and the coupling term gets stronger. However, even for slightly colored regions ($C_{abc}\ll$) the prevention of the caustic becomes plausible in highly curved conditions since the Riemann tensor enters the rhs of relation (\ref{caupr}).

  We can extract further information about the last term of relation (\ref{simpRch}) if we decompose the curvature tensor. In GR the curvature tensor decomposes to a purely antisymmetric part (Weyl tensor), to a symmetric second rank tensor (Ricci curvature) and to a scalar part. Although in Finsler geometry the Riemann curvature is well defined, to the best of our knowledge a similar meaningful decomposition cannot be retrieved since $R_{abcd}$ does not possess the same symmetries.
  Nevertheless, we can impose particular symmetries for the curvature to obtain further insight in the same lines with the Riemannian case. In analogy with GR, the relation which characterizes an isotropic point in space-time is
  \begin{equation}
  l^dR_{dabc}=\mathfrak R(l_bg_{ac}-l_cg_{ab})
  \label{seccurv}
  \end{equation}
  where $\mathfrak{R}$  corresponds to the Ricci scalar in the Riemannian limit \cite{RundFB}. This limit is widely used in isotropic scenarios of gravitational collapse. Then, the condition (\ref{caupr}) takes the simplified form
  \begin{equation}
  \frac{1}{2}\kappa \rho+2\sigma^2<\mathfrak{R}lC_au^a,
  \end{equation}
where $C_a=C_{abc}g^{bc}$ is the Cartan torsion vector. Therefore, in the symmetric limit (\ref{seccurv}) only the time-like part of the element of support contributes to the avoidance of the caustic. Also, the focusing of the congruence can be avoided either for negative or positive $\mathfrak{R}$ depending on the orientation between Cartan's torsion vector and the
tangent to the flow lines.

\section{Osculating Riemannian FRW geometry }
Another method to explore some aspects of Finsler geometry is the osculating Riemannian approach. This limiting process restricts our analysis to a region of $F_n$ where the coordinate increments $y^a$ depend on the position $x^a$ (see for example \cite{RundFB,Oscul} and for the kinematics of deformable media \cite{cong}).  In that case relation (\ref{fmeten}) defines a purely Riemannian metric, namely
\begin{equation}
r_{ab}(x)=g_{ab}(x,y(x)),
\label{rab}
\end{equation}
and the imposed metricity implies the  Levi-Civita parallelism. Then, in our subregion we can use the Einstein field equations together with the energy-momentum conservation by following the usual arguments
\begin{equation}
 G_{ab}(x,y(x))=\kappa T_{ab}(x,y(x))\;\;\mbox{,} \;\; \nabla_bT^{ab}(x,y(x))=0\label{fiecon},
\end{equation}
where $G_{ab}=R_{ab}-{1\over2}Rr_{ab}$ stands for the Einstein tensor, while the nabla operator is the linear connection induced by the metric (\ref{rab}).

 Let us consider the interesting class of Randers spaces
\begin{equation}
F=\alpha+\beta,
\label{ran}
\end{equation}
with $\alpha=\sqrt{\alpha_{ab}y^ay^b}$ standing for a pseudo-Riemannian metric, while $\beta=b_a y^a$ is the Finslerian contribution. The vector field $b_a$ introduces a preferred direction in space-time as a phenomenological consequence of a Lorentz invariance violation. The metric function (\ref{ran}) interfaces a Riemannian space-time with Finsler geometry in a simple manner. The study of such space-times gives many new possibilities since it monitors departures from specific Riemannian examples.
 Concerning the process of gravitational collapse in general relativity the most characteristic example is given by a spherical symmetric and isotropic dust fluid. In that case, the geometry of space-time is well described by the FRW metric
\begin{equation}
\alpha_{ab}=\mbox{diag}(1,-\frac{a^2(t)}{1-Kr^2},-a^2(t)r^2,-a^2(t)r^2\sin^2\theta).
\label{FRW}
\end{equation}
Note, that $K=0,\pm1$ is the spatial curvature for  a flat, closed and open model respectively. This perfect dust-like configuration of the Riemannian limit can be transplanted in a Finslerian set up by inserting the metric tensor (\ref{FRW}) in relation (\ref{ran}). Let us assume the common Finsler case where the vector field $y^a$ is identified as the velocity of the congruence in question. Then, the 4-velocity $u_a$ of the fluid flow stands for the anisotropic variables $y^a$. By virtue of relation (\ref{ran}) the vector field $b_a$ is written into its irreducible parts
\begin{equation}
b_a=\beta u_a+\mathfrak{b}_a,
\end{equation}
where  $\mathfrak{b}_a=h_a^{\;\;b}b_{\,b}$ stands for a preferred direction in the instantaneous rest space of the fundamental observer.

 In order to keep close to the FRW symmetry we impose homogeneity and isotropy. This, assumption switches off the spatial part of $b_a$. Also, our analysis is restricted to monotonically decreasing functions of $\beta(a)$ to recover the FRW limit for large values of the scale factor. We consider the particular profile, $\beta=c_1a^{-n}$, where $c_1$ stands for a small constant. However, the following analysis holds for a large class of profiles. Projecting along the 4-velocity of the comoving observer relations (\ref{fiecon}) give back the evolution law for the scale factor
 \begin{equation}
 {1\over2}\dot{a}^2-{1\over6}\kappa\rho a^2(1+4\beta)+K\beta=-\frac{1}{2}K,
 \label{Frie}
 \end{equation}
together with the linear independent continuity equation
\begin{equation}
\dot{\rho}+\left(3\frac{\dot{a}}{a}+\frac{2\dot{\beta}}{1+2\beta}\right)\rho=0,
\label{cont}
\end{equation}
where $\beta$ remains an undetermined function of time.

\begin{figure}
\begin{center}
\includegraphics[width=0.48 \textwidth, angle=0]{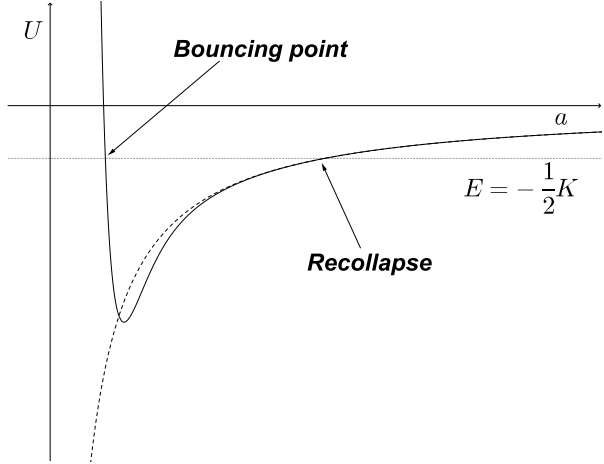}
\end{center}
\caption{\small A sketch for the mechanical analogue of a test-particle moving in the potential (\ref{pot}) for a closed spatial geometry with $\beta>0$ and $n>1$. The straight line represents the total energy, the dash line is the standard FRW potential and the solid one stands for the modified case (\ref{pot}). An initially contracting region starting from the right of  $U_{min}$ will accelerate  and then decelerate until the bouncing point. After the bounce, it enters an accelerating expanding phase until it crosses $U_{min}$. Then, the region decelerates until the self-gravity of the fluid dominates. The dust-ball recollapses and the process starts over revealing an oscillatory behavior for the 3D-volume element.     }
\label{bounce}
\end{figure}

The Newtonian kinematical analogue of relation (\ref{Frie}) is the motion of a point particle with kinetic energy $K={1\over2}\dot a^2$, and   total energy $E=-\frac{1}{2}K$. We can directly integrate the first order differential equation (\ref{cont}) to obtain the solution for the energy density
 \begin{equation}
 \rho=\rho_0\frac{a^{-3}}{1+2\beta}.
 \end{equation}
 In that case, our hypothetical particle falls in the potential
\begin{equation}
 U=-{1\over6}\kappa\rho_0 a^{-1}\frac{1+4\beta}{1+2\beta}+K\beta\label{pot}.
 \end{equation}
 The modified FRW potential (\ref{pot}) corresponds to a collapsing dust-like fluid if the scale factor $a$ is a decreasing function of time. On the other hand, it is natural to expect that the Finslerian contribution fades out as the fluid expands. Then, to recover the FRW behavior for large values of the scale factor we must impose that $\beta$ decrease faster than $a^{-1}$ ($n>1$ for our profile).   In that case, for late times of the collapsing process the $\beta$ contribution in the rhs of (\ref{pot}) can play the role of a gravitational repeller. The main effect comes from the coupling term with the spatial curvature and depends on the Riemannian 3D-geometry ($K=0,1,-1$). In the same line with Raychaudhuri's equation (\ref{Raych1}), we retrieve again a color-curvature contribution since $\beta$ is the quantity that breaks the quadratic restriction (\ref{ran}). Moreover, to resist gravitational pull of matter the spatial curvature $K$ and the Finslerian contribution $\beta$ must be of the same sign. Notice, that the main effect fades out in case of  flat spatial section.

Consider a contracting region of our hypothetical space-time where initially the scale factor lies in the standard FRW region of the potential (\ref{pot}).  Due to the self attraction of matter  the contraction will be in an accelerating phase as in standard gravitational physics.  As the region shrinks, the coupling term between the spatial curvature and color starts to dominate and acts as an effective pressure against the pull. Eventually, it forces the collapse to a decelerating phase until the fluid bounces to the potential at the turning point
 \begin{equation}
 a_{b}\approx\frac{\kappa\rho_0}{3(1+2\beta_b)},
 \end{equation}
 where the subscript $b$ denotes the value at the bouncing point.
 After the bounce the medium expands in an accelerating manner that turns to a decelerating phase until the gravity dominates again. Then, the fluid recollapse and the process repeats itself to infinity revealing an oscillatory behavior for the scale factor (see figure-\ref{bounce}). A similar behavior is retrieved in the context of very special relativity  for the Bogoslovsky's line element \cite{imper} and on tangent Lorentz bundles \cite{lorbund}. The bounce is also possible in an open model if $\beta<0$ but an expanding phase will never recollapse. Finally, in  case of flat spatial sections the singular fate of a collapsing dust-like fluid is unavoidable.

\section{Discussion and conclusions}
In summary, the intrinsic anisotropy imposed by the non-quadratic metric function $F(x,y)$ gives birth to various kinematical complications. The evolution of the deviation vector along the medium's flow lines monitors the internal deformation and involves up to second order terms of $\xi^a$. Combined with the three distinct Ricci identities (\ref{Ric1})-(\ref{Ric3}) that we get from the two types of covariant derivative, the evolution equations for the expansion, shear and vorticity of the flow seems difficult to be given in closed form expressions. At this point one can either treat the tangent bundle as his base manifold and split his tensorial quantities to vertical and horizontal components (for textbook treatment see \cite{Miron}),  or proceed to further geometric assumptions to get some analytical or numerical results. The first approach gives analogue relations with GR in the horizontal and vertical sub-bundles of $\mathcal{TTM}$ providing a way to study focusing of congruences in the total space of the manifold \cite{Stavrin}. However, one has to ``translate'' the covariant $\mathcal{TTM}$-expressions back to the actual deformation of a medium's cross-section.

 On the other hand, at a first ``brute-force'' attempt we may try to solve the global problem for the congruence and its connecting vector for particular geometric conditions. Then, we can track back the deformation of a cross section through the evolution of the deviation vector with respect to the arclength. The latter is given by the absolute differentiation along medium's flow lines and involves two distinct covariant derivatives. The first one is along the tangent to the flow lines $u^a={dx^a\over ds}$, while the other is along the evolution direction of the element of support, $\dot{\ell}^a={Dl^a\over ds}$.   In the main bulk of the article we restrict our analysis for simplicity to normalized time-like congruences that depend solely in the position coordinates, $u^a=u^a(x)$. This particular class of Finslerian integrable curves is widely used  to retrieve simpler formulas and isolate particular properties of the manifold \cite{RundFB}. They are closely related to the Riemannian ones and we can Lie drag a tensor field along them in a simple and straightforward way. However, the $u^a$-bundle still ``feels'' the color of the structure mainly through the Cartan tensor. In fact, this is clearly depicted in the constrain relation for the Killing vectors (\ref{Killng}) where the bundle can anisotropically expand in contrast to GR.

  The Lie derivative plays a central roll since  we define the deviation vector as a Lie transported spatial congruence along the flow, $\pounds_u\xi_a=0$  \cite{schutz}. Even in case of a deviation vector that depends solely in the position the evolution of a cross-section involves up to second order terms with respect to $\xi^a$. For a given direction $l^a$, the first order deformation is given by the horizontal derivative of $u^a$ and the second  order by the Cartan tensor (\ref{DxiTor}). The example of an isotropically evolving volume element is given where the expansion of the flow cannot be determined \emph{ab initio}. In a simplistic scenario we solve for the volume of the congruence and demonstrate how an exponentially expanding region deflates or reach a caustic due to the Finslerian contribution. This suggests that emergent effects of QG may have a crucial kinematic reaction to inflationary models of the early universe and deserves further investigation.

   Possible GR modifications are more likely to appear in highly curved regions where  QG physics may emerge. In GR, such conditions  favor the formation of a singularity and/or a caustic of the congruence. Hence, the breaking of the local symmetry of GR opens up the possibility to evade the focusing of neighboring particles and the space-time singularities (see for example \cite{QGraych,Gasperini}). From a kinematical perspective, any modification of the curvature theory will directly affect the deformation of the $u^a$-bundle \cite{QGraych}. That is the case also in Finsler geometry where the curvature theory is extended with the introduction of color. The later monitors departures from the Lorentz symmetry since it reflects the breaking of the quadratic restriction on the distance module. Hence, in Finsler geometry the internal motion of a time-like flow is affected by the color (LV) and its variations.

   We concentrate our analysis in the expansion of a time-like flow. The complexity of the general problem forces us to pick a particular class of congruences along which the preferred direction imposed by the element of support is parallel displaced. Then, for a position dependent time-like bundle Raychaudhuri's formula can be retrieved. In that case, general conclusions for the expansion dynamics can be derived. The more interesting Finslerian effect is a coupling between curvature and color. This term is more likely to become dominant in high curvature environments; exactly where we speculate that QG effects may become important. Under certain conditions it can reproduce a repulsive effect to neighboring particles or add to the gravitational pull. The focusing theorem for our restricted congruence is given where irrotational geodesics of dust like matter can avoid the caustic. Nevertheless, the caustic prevention does not guarantee that the real space-time singularity vanish. However, it points out a possible QG mechanism that can slow down or even stop gravitational collapse. The formalism can be extended to study the effect on the shear and vorticity of the medium. In the same context the covariant perturbations over an almost FRW model can give back some imprints of color in the spatial distribution of the cosmological fluid \cite{Relcos}. Also, the Raychaudhuri's equation for null geodesics \cite{null} would be desirable since it is of central importance in gravitational lensing and distance measurements \cite{Perlick:2004tq}.

   One of the most famous solutions of Einstein's theory of gravity is the FRW space-time. It describes an isotropic and homogeneous fluid and it is a good approximation for the cosmological medium and for the interior of a star. In GR it clearly monitors the gravitational pull of matter through the evolution of the scale factor. The contracting phase is always accelerating and leads to the Big-Bang singularity, $a\rightarrow0$. In phenomenological models of quantum gravity and for non-standard matter fields the Big-Bang singularity is avoided in various examples \cite{Novello:2008ra}. Usually, the medium before it reaches the singular point bounces back and enters an expanding phase. This models are candidate alternatives to inflation and they can also prevent the formation of the black hole horizon \cite{farw}. We recover the same behavior for an almost FRW model using the Randers metric in the osculating Riemannian limit. Moreover, the bounce occurs only when the space-time is spatially curved. The mechanism that prohibits the formation of the singularity is a  coupling between the spatial curvature and the variable that breaks the Lorentz symmetry. This result is in alliance with Raychaudhuri's equation and reveals from a geometric perspective that departures from the local symmetries of GR may lead to non-singular space-times.

\begin{acknowledgments}

The authors would like to thank the University of Athens (Special Accounts for Research Grants) for the support to this work. We are also grateful to C.G.Tsagas and S.Vacaru for the helpful discussions.

\end{acknowledgments}

\end{document}